\documentclass[10pt,a4paper, twocolumn, aps, pra, showpacs, longbibliography]{revtex4-1}
\usepackage[utf8x]{inputenc}
\usepackage{ucs}
\usepackage{amsmath}
\usepackage{amsfonts}
\usepackage{amssymb}
\usepackage{makeidx}
\usepackage{cellspace,booktabs}
\usepackage{natbib}
\usepackage{lipsum}
\usepackage{bm}
\usepackage{hyperref}

\usepackage{graphicx}
\usepackage{braket}
\usepackage{microtype}

\newcommand{\mat}[1]{\ensuremath{\mathbf#1}} 
\newcommand{\nl}{nonlinear}
\newcommand{\trans}{^{\hspace{-0.5pt}\text{T}}}

\renewcommand{\dag}{^{\dagger}}

\begin{document}
\title{Normal modes of a superconducting transmission-line resonator with embedded lumped element circuit components}
\author{Henrik Lund Mortensen}
\affiliation{Department of Physics and Astronomy, Aarhus University, DK-8000 Aarhus C, Denmark}
\author{Klaus Mølmer}
\affiliation{Department of Physics and Astronomy, Aarhus University, DK-8000 Aarhus C, Denmark}
\author{Christian Kraglund Andersen}
\thanks{Email: ctc@phys.au.dk}
\affiliation{Department of Physics and Astronomy, Aarhus University, DK-8000 Aarhus C, Denmark}

\begin{abstract}
We present a method to identify the coupled, normal modes of a superconducting transmission-line with an embedded lumped element circuit. We evaluate the effective transmission-line non-linearities in the case of Kerr-like Josephson interactions in the circuit and in the case where the embedded circuit constitutes a qubit degree of freedom, which is Rabi coupled to the field in the transmission-line. Our theory quantitatively accounts for the very high and positive Kerr non-linearities observed in a recent experiment [\mbox{M. Rehák \emph{et.al.}}, Appl. Phys. Lett. \textbf{104}, 162604], and we can evaluate the accomplishments of modified versions of the experimental circuit.
\end{abstract}
\maketitle

\section{Introduction}
The quantum mechanical interaction between light and matter is of major importance in non-linear optics and quantum optics. Recently, the field of circuit QED \cite{wallraff2004strong, PhysRevA.69.062320} has seen immense progress with superconducting circuits playing the role of artificial atoms which can interact with confined microwave photons and produce quantum optical effects such as vacuum Rabi-splitting \cite{wallraff2004strong, abdumalikov2008vacuum}, parametric amplification \cite{castellanos2008amplification, PhysRevB.83.134501, rehak2014parametric}, single atom lasing \cite{astafiev2007single, PhysRevB.91.104516}, and photon-mediated interactions \cite{van2013photon}. The photonic non-linearity of these experiments arises from the non-linear interactions in Josephson junctions in the superconducting circuits which are embedded in microwave transmission-line waveguides and resonator architectures.

Linear, classical microwave circuits can be fully characterized by their impedance \cite{pozar2009microwave} which, e.g., reveals their transmission properties and resonances. The quantum description follows in a similar manner \cite{PhysRevLett.108.240502}, and a transmission-line resonator can thus be described by its classical eigenmodes which directly translate into a quantum Hamiltonian as a sum of harmonic oscillators. Josephson junctions, on the other hand, are non-linear elements, i.e., they are not described as harmonic oscillators and their incorporation in circuit architectures causes non-linear and potentially non-classical quantum effects. A single Josephson junction is described by an energy potential which is a cosine function of the quantum mechanical phase variable. If the cosine potential is deep and supports many eigenstates, we can expand the potential and obtain an effective Kerr-effect in our system \cite{PhysRevA.86.013814, leib2012networks, eichler2014controlling, PhysRevLett.108.240502}, while if it supports only few eigenstates with irregular level spacing, the dynamics may be restricted to the  two lowest eigenstates which constitute a single qubit \cite{nakamura1999coherent}. By shunting the Josephson junction with a large capacitor and coupling it capacitively to a resonator or by adding several Josephson junctions in a superconducting loop, one can obtain the so-called transmon qubits \cite{PhysRevA.76.042319} or flux qubits \cite{mooij1999josephson, yan2015flux}.

\begin{figure}[b]
  \centering
  \includegraphics[width=240pt]{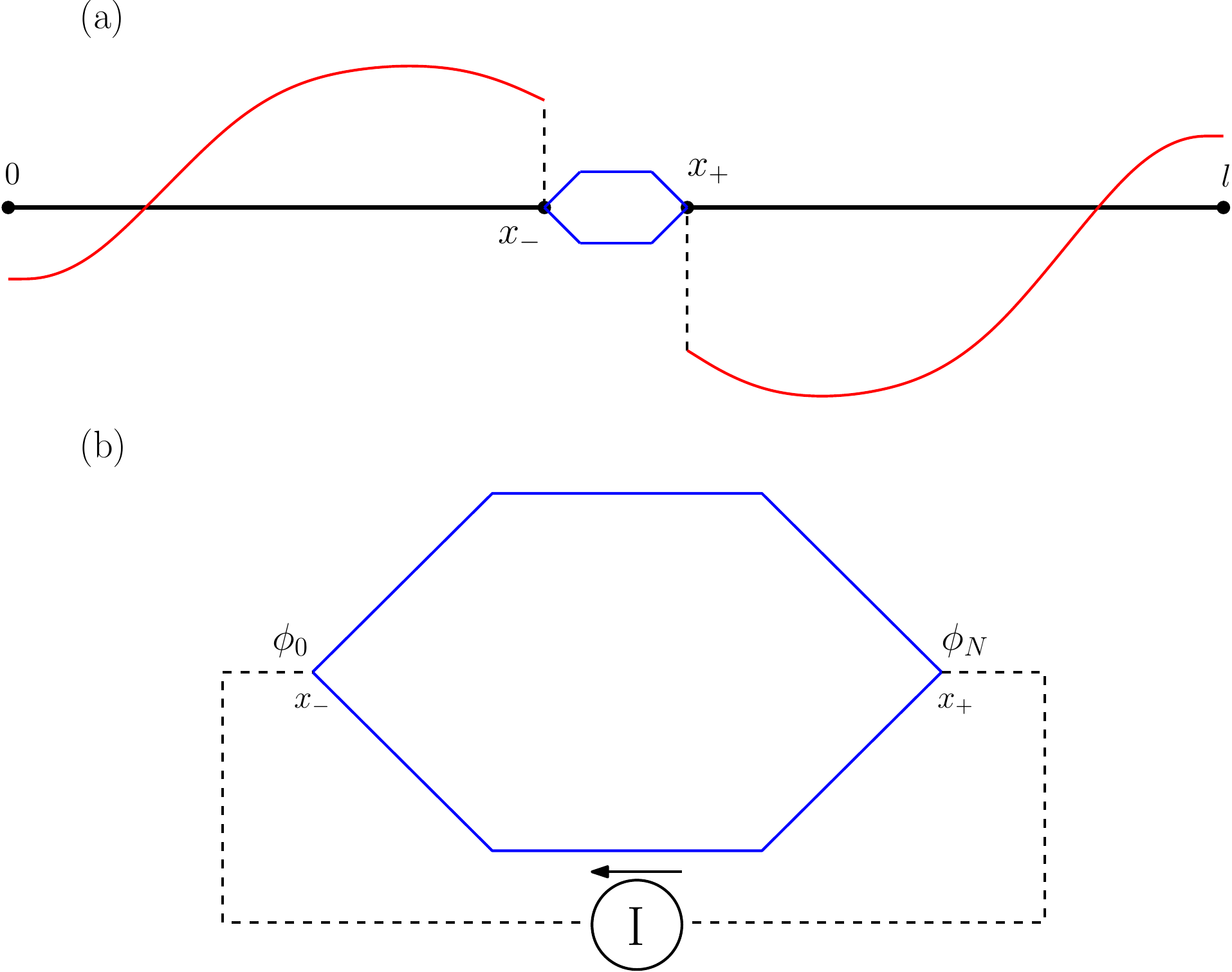}
  \caption{(a) Schematic drawing of an inline lumped element circuit (blue) which introduces a boundary value discontinuity in the mode envelopes of a transmission-line (red) and hence perturbs both the eigenfrequencies and the coupling strengths compared to the bare mode case. (b) The transmission-line is modeled as a current source to the inner circuit, with the current given by the transmission-line flux variable}
  \label{fig: resonator with normal mode}
\end{figure}

As illustrated in Fig. \ref{fig: resonator with normal mode}~(a), when small quantum circuit systems are embedded in waveguide resonators, they enforce local boundary conditions on the transmission-line variables, and they may hence not only interact with but also significantly modify the waveguide mode structure. In this work we will present a general approach to describe the linear eigenmodes of an arbitrary circuit embedded in a transmission-line resonator. The eigenmodes of the coupled systems are then taken as the basis for the pertubative inclusion of non-linear interactions. Our approach builds on the Lagrangian formalism \cite{devoret, PhysRevA.29.1419} and it generalizes prior approaches used to describe weakly non-linear systems \cite{PhysRevA.86.013814, PhysRevB.92.104508, leib2012networks, PhysRevLett.112.223603} applied, e.g., as microwave amplifiers  \cite{eichler2014controlling, PhysRevLett.113.110502}. Motivated by a recent experiment \cite{rehak2014parametric} that used a pair of flux qubits inside a resonator for parametric amplification, we will extend our formalism to specifically deal with flux qubits in the system.

The article is organized as follows: In Sec. \ref{sec:continuous} we review the general spanning tree method to deal with lumped element circuit and with waveguide components, and we identify the normal modes of a resonator with an arbitrary embedded circuit and their quantum mechanical and non-linear interactions. Section \ref{sec:fluxqb} deals with the inclusion of a strongly coupled flux qubits which causes Rabi splitting of the mode structure and which cannot be described directly by a perturbative Kerr-interaction. In Sec. \ref{sec:experiment} we apply our formalism to a recent experiment and in Sec. \ref{sec:conc}, we conclude and provide an outlook.

\section{Coupling a discrete circuit with a continuous mode}
\label{sec:continuous}
Standard techniques exists to quantize the modes of a resonator \cite{PhysRevA.69.062320} and the degrees of freedom of coupled, lumped element superconducting circuits \cite{devoret}. The combination of a discrete circuit and a resonator mode has also been considered in special cases \cite{PhysRevA.69.062320, PhysRevA.86.013814, PhysRevA.91.023828}, but without establishing a general framework. In this section, we use the standard nodal technique to provide such a general, practical theory as a generalization of the method in Ref. \cite{PhysRevA.86.013814}.

\begin{figure}[t]
\includegraphics[width=\linewidth]{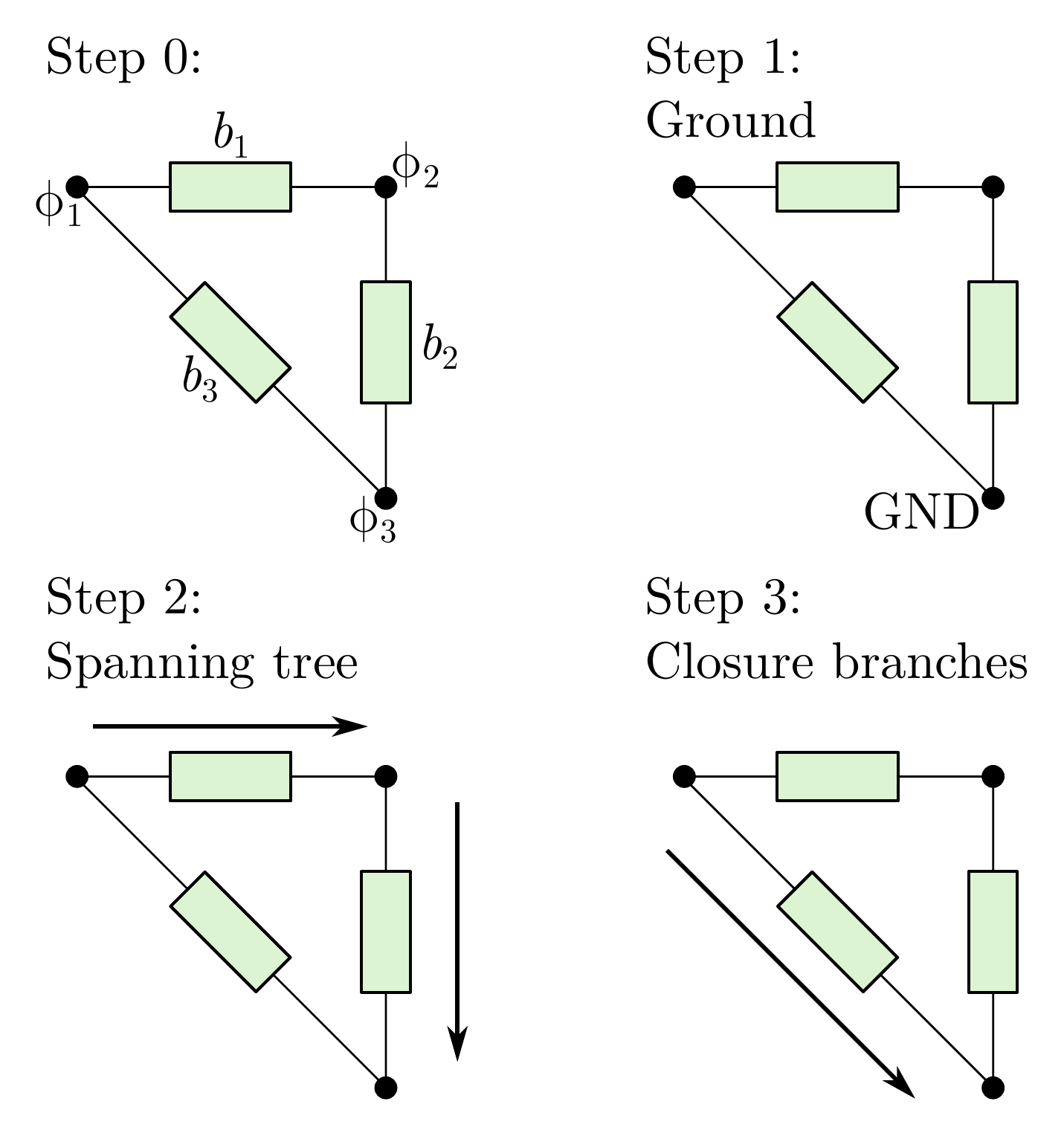}
\caption{The steps required to analyze an arbitrary network. For illustrative purposes we have in step 0 three nodes, $\phi_n$, connected by the branches $b_i$. The branches can be either inductive or capacitive elements. The first step is to define a ground node, here the node $\phi_3$. Then in step 2, we choose a spanning tree, $b_1$ and $b_2$, with a given direction, which defines a positive direction of a voltage drop. Then finally we identify the closure branches and their direction, here only $b_3$.} \label{fig:network}
\end{figure}

Well-suited coordinates for a Lagrangian description of circuit electron dynamics \cite{devoret} are the so-called node flux variables $\phi_n(t)$, given as the time-integral of the voltage at different nodes in the circuit. A complete, consistent description of all $N$ nodes of a circuit is given by considering the inductor and capacitor elements connecting the nodes as branches and choosing a spanning tree of paths such that every node is reached from a chosen reference (ground) node through only one path. The rest of the branches are then referred to as closure branches. This approach is sketched in Fig.~\ref{fig:network} and following these steps takes care of the gauge degree of freedom in analyzing electrical circuits \cite{devoret}. Let $\delta_m$ denote the time-integral of the voltage difference, the branch flux, across each branch element. The flux variables $\phi_n(t)$ are then given by the sum of the branch fluxes along the path connecting the nodes to ground, and it can, in turn be written as a sum over all branch fluxes
\begin{equation}
  \label{eq. node flux}
  \phi_n(t) = \sum_b S_{nb}\delta_b(t),
\end{equation}
 where the factors $S_{nb}=\pm 1$ if the branch $b$ is in the path towards the node $n$, with the sign $+(-)$ depending on the voltage difference being defined towards (away from) the node, and $S_{nb}=0$ if $b$ is not in the path connecting the node $n$ to the ground node. With the node description in place, we obtain the transformation from node to branch variables coming from all $S_{nb}$,
\begin{equation}
  \label{eq. node to branch}
  {\delta}(t) = \mat{K}\vec{\phi}(t),
\end{equation}
where $\mat{K}$ is the the incidence matrix,
\begin{equation}\label{eq. incidence matrix constructions}
\mat{K}_{ij} = \Bigg\{
     \begin{array}{lr}
       1 &  \text{if branch $i$ has node $j$ at the + end}\\
       -1 &  \text{if branch $i$ has node $j$ at the - end}\\
       0 & \text{otherwise},
     \end{array}
\end{equation}
and the variables $\delta(t)$ and $\vec{\phi}(t)$ are vectors containing all $\delta_n$ and $\phi_n$, respectively. We remark that, the matrix $K$ is uniquely defined once the steps of Fig.~\ref{fig:network} are followed.

For a circuit whose branches are only inductors and capacitors we define the following matrices in the basis of branch variables,
\begin{align}
  \label{eq. L and C mat}
  \mat{L} &= \text{diag}\{ 1/L_1, \dots, 1/L_M\} \\
  \mat{C} &= \text{diag}\{ C_1, \dots, C_M\},
\end{align}
where $L_i$ and $C_i$ are the inductance and capacitance of branch $i$ (replace $1/L_i$ with 0 if the inductance of the $i$th branch is zero). The Lagrangian description of the circuit is obtained by treating inductive energy, $E_L = \delta/2L$, and capacitive energy, $E_C = C\partial_t \delta/2$, as potential and kinetic energy respectively, and using Eq.~\eqref{eq. node to branch} to transform from flux to branch variables,
\begin{equation}
  \label{eq. Lagrangian lumped element}
  \mathcal{L}_{lin} \left( \vec{\phi}(t), \partial_t\vec{\phi}(t) \right) = \frac{1}{2}\partial_t\vec{\phi}(t)\trans{}\tilde{\mat{C}}\hspace{1pt}\partial_t\vec{\phi}(t) + \frac{1}{2}\vec{\phi}(t)\trans{}\tilde{\mat{L}}\vec{\phi}(t),
\end{equation}
with $\tilde{\mat{C}} = \mat{K}\trans{} \mat{C} \mat{K}$ and $\tilde{\mat{L}} = \mat{K}\trans{} \mat{L} \mat{K}$.
This Lagrangian yields the equations of motion for the flux variables
\begin{equation}
  \label{eq. equation of motion}
  \tilde{\mat{C}}\partial_t^2{\vec{\phi}}(t) +  \tilde{\mat{L}}\vec{\phi}(t)= 0,
\end{equation}
which is automatically equivalent the Kirchoff equations for the circuit \cite{devoret}.

With the above giving an appropriate description for a discrete lumped element circuit, we need also to consider distributed elements such as a transmission-line resonator as our goal is to combine the two. A transmission-line can be considered as a series of LC-circuits of length $\Delta x$. In the limit $\Delta x \rightarrow 0$, the flux at each node becomes a continuous function, which we for convenience will denote $\psi(x,t)$, but it plays the same role as $\vec{\phi}(t)$ in the discrete case. Treating the transmission-line first as a lumped element circuit, we can obtain an equation of motion on the form of Eq. \eqref{eq. equation of motion}. For a series of $LC$-circuits the corresponding matrix $K$ connects each node with both neighboring nodes which generates inductive energy leading to terms of the form $(2\psi_n - \psi_{n+1} - \psi_{n-1})/\Delta x^2$ in the equation of motion. This is equivalent to the curvature of a continuous function when taking the continuum limit and the equation of motion emerges as the wave equation,
\begin{equation}
  \label{eq. wave equation}
  \partial_{t}^2\psi(x,t) - v^2\partial_x^2\psi(x,t) = 0,
\end{equation}
with $v = \sqrt{1/L_0C_0}$ with $L_0$ and $C_0$ being the inductance and capacitance per length of the transmission-line. Notice, that the current in the transmission-line is given by
\begin{equation}
  \label{eq. flux-current relation}
I(x,t) = -\frac{1}{L_0} \partial_x \psi(x,t).
\end{equation}

\subsection{Normal modes of the combined system, linear theory}
For a finite transmission-line resonator embedded with a linear lumped element circuit, the dynamics of the combined system can be described in terms of common normal modes for the full system. Such normal modes consist of both a continuous and a discrete part which we now will denote $\psi_m(x,t)$ and $\vec{\phi}_m (t)$ respectively and both are oscillating at a common angular frequency $\omega_m$. However, the functions, $\psi_m$ and $\vec{\phi}_m$, depend on $\omega_m$ and must by found by solving one combined eigenvalue problem including both the discrete and the continuous part. The aim of this section is therefore to formulate such an eigenvalue problem.

For the frequencies of interest, the spatial solutions of the continuous wave equation along the transmission-line vary on the length scale $l$ of centimeters, while the lumped circuits are much smaller and described by discrete flux variables. The inline circuit thus introduces an effectively discontinuous drop in the transmission-line flux, $\Delta \psi$, across the embedded circuit \cite{PhysRevA.86.013814} (see Fig.~\ref{fig: resonator with normal mode} (a)). For each normal mode, we therefore define
\begin{equation}
  \label{eq. flux drop}
  \Delta \psi_m(t) = \psi_m(x_+,t) - \psi_m(x_-,t).
\end{equation}
The spatial transmission-line components of the system eigenmodes further satisfy boundary conditions at $x = 0$ and $x = l$, while accommodating also the flux drop at $x = x_\pm$.

To calculate $\Delta \psi_m(t)$, we treat the transmission-line as a current source for the inline circuit, see Fig.~\ref{fig: resonator with normal mode} (b). Therefore to diagonalize the system and thus find $\psi_m$ and $\vec{\phi}_m$ and their appropriate $\omega_m$, the Euler-Lagrange equation, Eq. \eqref{eq. equation of motion}, with an extra current drive must be solved together with wave-equation for the transmission line resonator. Restricted to the discrete flux variables, the eigenmode equations are thus given by
\begin{equation}
  \label{eq. of motion normal}
  (\tilde{\mat{L}} - \omega_m^2\tilde{\mat{C}})\phi_m(t) = \vec{I}_m(t).
\end{equation}
with the vector of driving currents,
\begin{equation}
  \label{eq. f_m}
\vec{I}_m(t) =
\begin{pmatrix}
    -I_0(t)\\
    0\\
    \vdots\\
    0\\
    I_N(t)
\end{pmatrix}.
\end{equation}
where the two current terms represent the resonator mode driving of the node 0 and the node $N$ flux variables. The eigenmode equation, Eq. \eqref{eq. of motion normal}, has a pole at the eigenfrequencies of the inner circuit, but since we are interested in the dressed resonator modes this is not a problem. Therefore, provided the eigenfrequency $\omega_m$ is not an eigenfrequency of the isolated inline circuit, the inhomogeneous equation, Eq. \eqref{eq. of motion normal}, is solved by inversion,
\begin{equation}
  \label{eq. solution for phi}
  \vec{\phi}_m(t) = \text{inv}(\tilde{\mat{L}} - \omega_m^2\tilde{\mat{C}}) \cdot \vec{I}_m(t).
\end{equation}
Now, we can apply the relation between the currents and the derivative of the transmission-line flux $I_0 = \partial_x \psi(x_-) / L_0 = I_N = \partial_x \psi(x_+) / L_0$. The transmission-line flux drop across the circuit is given by the flux at the 0th and $N$th node (at $x_-$ and $x_+$), therefore, combining Eq.~\eqref{eq. solution for phi} with Eq.~\eqref{eq. flux drop} we obtain the equation
\begin{align}
  \label{eq. trancendental equation for omega}
  \Delta \psi_m(t) & = \text{inv}(\tilde{\mat{L}} - \omega_m^2\tilde{\mat{C}})_{0,0} \hspace{1pt}  I_0(t) \nonumber \\
  & \quad -  \text{inv}(\tilde{\mat{L}} - \omega_m^2\tilde{\mat{C}})_{0,N} \hspace{1pt}  I_N(t) \nonumber \\
  & \quad -  \text{inv}(\tilde{\mat{L}} - \omega_m^2\tilde{\mat{C}})_{N,0} \hspace{1pt}  I_0(t) \nonumber \\
  &\quad +   \text{inv}(\tilde{\mat{L}} - \omega_m^2\tilde{\mat{C}})_{N,N} \hspace{1pt}  I_N(t).
\end{align}
As alluded to earlier, this equation must be solved together with the wave equation, Eq. \eqref{eq. wave equation}, which has the normal mode solutions $\psi_m(x) \propto \cos (k_{\omega_m} x + \phi_m^{\pm})$. Here we have $k_{\omega_m} \propto \omega_m$ as the wave number and $\phi_m^{\pm}$ the flux set by the boundary conditions at $0$ and $l$, where we use $\pm$ to note the flux for $x$ smaller or larger than $x_\pm$. Therefore, the self-consistent solution of the wave equation solution and Eq. \eqref{eq. trancendental equation for omega} yield a transcendental equation for $\omega_m$.

The combined continuous and discrete modes are orthogonal under the energy inner products \cite{taylor},
\begin{align}
  \label{eq. energy inner products C}
  \braket{\{ \psi_m,\phi_m\},\{ \psi_n,\phi_n\} }_C &= C_0\int_{\text{TL}}dx \psi_m(x)\psi_n(x) \nonumber\\
  &\quad+ \phi_m\trans{}\tilde{C}\phi_n \nonumber\\
  &= C_\Sigma\delta_{nm} \\
 \braket{\{ \psi_m,\phi_m\},\{ \psi_n,\phi_n\} }_L &= \frac{1}{L_0}\int_{\text{TL}}dx \partial_x\psi_m(x)\partial_x\psi_n(x) \nonumber\\
  &  \quad+ \vec{\phi}_m\trans{}\tilde{L}\vec{\phi}_n \nonumber\\
  &= \frac{1}{L_\Sigma^m}\delta_{nm}.  \label{eq. energy inner products L}
\end{align}
Equation (\ref{eq. energy inner products C}) and Eq. (\ref{eq. energy inner products L}) enable the diagonalization of the full Lagrangian, given by Eq. \eqref{eq. Lagrangian lumped element} and $\mathcal{L}_{tl} = \int_0^l \frac{C_0}{2} (\partial_t \psi)^2 + \frac{1}{2L_0} (\partial_x \psi)^2 dx$. Fixing the normalization of $\psi_m$ and $\phi_m$ by Eq. \eqref{eq. energy inner products C} we obtain $\omega_m = 1/\sqrt{C_\Sigma^{\phantom{m}} L_\Sigma^m}$.

\subsection{Quantization of the system}

With the normal modes obtained by solving Eq. \eqref{eq. trancendental equation for omega} together with Eq. \eqref{eq. wave equation} and \eqref{eq. solution for phi}, we write a harmonically varying eigensolution as
\begin{align}
\varphi_m (x,t) = \varphi_m(t) u_m(x)
\end{align}
with $x \in \lbrace [0;l], 0 ,1, \ldots , N \rbrace$. Here, the stationary mode function $u_m(x)$
\begin{align}
u_m(x) = \begin{cases} \psi_m(x) & \text{ if } x\in[0;l] \\ \vec{e}_x^{\;\text{T}} \vec{\phi}_m & \text{ if } x \in \lbrace 0,1, \ldots N \rbrace \end{cases}
\end{align}
with $\vec{e}_x$ being the $x$th unit vector, satisfies the normalization given by Eq. \eqref{eq. energy inner products C}. $\varphi_m(t)$ is the time dependent amplitude with the canonically conjugate variable $q_m = \partial \mathcal{L} / \partial (\partial_t \varphi_m)$. 

We quantize the system by imposing the operator commutator $[\hat{\varphi}_m, \hat{q}_m] = i\hbar$.  For convenience, we introduce the annihilation and creation operators $a_m$ and $a_m\dag$ as
\begin{align}
\hat{\varphi}_m = \sqrt{ \frac{\hbar}{2C_\Sigma \omega_m} } (a_m\dag + a_m), \label{eq:phim} \\
\hat{q}_m = i\sqrt{ \frac{\hbar}{2C_\Sigma \omega_m} } (a_m\dag - a_m). \label{eq:qm}
\end{align}
and we transform the Lagrangian into a sum of harmonic oscillator Hamiltonians
\begin{align}
H_0 = \sum_m \hbar \omega_m a_m\dag a_m^{\phantom{\dagger}}.
\end{align}

\subsection{Josephson non-linearity}

So far, we considered systems containing only linear elements such as capacitors and inductors, leading to harmonic eigenmodes. These solutions are modified and the dynamics changes when Josephson junctions are introduced, with the characteristic anharmonic Josephson energy,
\begin{align}
  E_J(\delta(t))  &= E_J \cos \left( \frac{2\pi}{\Phi_0}\delta(t) \right) \\
  &= E_J - \frac{\delta(t)^2}{L_J} + \mathcal{O}(\delta(t)^4)  \label{eq. Josephson energy}
\end{align}
where $1/L_J = (2\pi/\Phi_0)^2 E_J$ sets the Josephson inductance, $\delta(t)$ is the flux drop across the junction and $\Phi_0$ the magnetic flux quantum. For the treatment in this work, $\delta(t)$ will be proportional with the phase drop over the full inner circuit, $\Delta \psi$, and is calculated from $\vec{\phi}$. If the Josephson non-linearity is weak, the linear behaviour of the Josephson Junction can modelled by a capacitor and inductor in parallel and the expansion in the second line of \eqref{eq. Josephson energy} can be directly incorporated in the harmonic oscillator eigenmode description. In this section, we proceed and consider the next term in Eq. \eqref{eq. Josephson energy}, yielding a higher order correction to the Hamiltonian
\begin{align}
H_{\text{nl}} = -\hspace{-5pt}\sum_{i\in \lbrace JJ \rbrace} \frac{2\pi^4}{3\Phi_0^4} E_{J,i} (\hat{\delta}_i)^4,
\end{align}
where $\hat{\delta}_i$ is set as in Eq. \eqref{eq. node to branch} and the summation is over all Josephson junctions in the circuit embedded in the  transmission-line. This way of re-introducing the higher order terms is only valid if the energy associated with the non-linear part is much smaller than $\hbar \omega_m$~\cite{PhysRevA.86.013814, PhysRevLett.108.240502}. Expanding the operator expression on the eigenmode creation and annihilaton operators,cf.,  Eq. \eqref{eq:phim}, and keeping only the energy preserving Kerr-like terms, we obtain the Hamiltonian
\begin{align}
H_{\text{nl}} = \sum_{m,n} \frac{\hbar K_{mn}}{2} (a_m\dag a_m a_n\dag a_n + a_m\dag a_m)
\end{align}
with
\begin{align}
K_{mm} = -\hspace{-5pt}\sum_{i\in\lbrace JJ \rbrace} \frac{\pi^4\hbar}{\Phi_0^4} \frac{E_{j,i}}{C_{\Sigma}^2 \omega_m^2} \Delta u_{m,i}^4 \label{eq:Kmm}
\end{align}
with $\Delta u_{m,i}$ denoting the difference in the normalized eigenmode function $u_m$ over the $i$th Josephson junction and similarly we find $K_{mn} = -2\sqrt{K_{nn}K_{mm}}$.

By finding the consistent eigenmodes and their eigenfrequencies, which may differ significantly from the bare transmission-line resonances, we have taken the linear couplings in the systems fully into account, and we have effectively treated the higher order non-linear terms as a perturbation. This perturbation affects the amplitude and population dynamics of the modes but not their spatiotemporal character. The Kerr-like field interaction \eqref{eq:Kmm} thus leads to a number of quantum optical effects, while cross-Kerr effects, via $K_{mn}$ represent dispersive interactions between the field and the circuit degrees of freedom or among the latter. In the following section, we shall deal with cases, where a different procedure is necessary because the non-linear coupling significantly affects the mode structure.

\section{Coupling to a flux qubit}
\label{sec:fluxqb}

For flux qubits, Josephson Junctions are contained within a closed loop in the circuit. In the linear regime, currents in such loops constitute additional internal modes that cannot be identified with Eq. \eqref{eq. trancendental equation for omega} and do not couple to the transmission-line, and they are thus not relevant for the dynamics of the normal modes found above. Due to the non-linearity of the Josephson elements, these modes may, however, couple strongly to other modes. In particular, if an embedded loop constitutes a qubit degree of freedom we may obtain a Jaynes-Cummings like coupling between the qubit and the transmission-line mode, which, in turn, can lead to a Rabi-splitting of the modes --  a highly \nl{} effect not captured by the perturbative Kerr terms. Therefore, we need to find the resonator modes using the method of Sec.~\ref{sec:continuous} and combine it with a proper description of the qubit degrees of freedom.

To illustrate this and provide a method to deal with the quantum dynamics of such systems, we consider a flux qubit consisting of a superconducting loop with three (or more) Josephson junctions. Following the treatment of Bourassa \emph{et.al} \cite{PhysRevA.80.032109} we write the Lagrangian for the flux qubit as
\begin{align}
\mathcal{L}_{fq} = \sum_{k=1}^3 \bigg[ \frac{C_{J,k}}{2} \dot{\phi}_k^2 + E_{J,k} \cos(2\pi \phi_k / \Phi_0) \bigg] \label{eq:fluxqblagrange}
\end{align}
with $k$ summing over the three junctions. The junctions 1 and 3 are identical with $C_{J,1} = C_{J,3} = C_J$ and $E_{J,1} = E_{J,3} = E_J$, while the center junction 2, referred to as the $\alpha$-junction, has $E_{J,2} = \alpha E_J$ and $C_{J,2} = \alpha C_J$ with $\alpha <1$. Note that the inner circuit may have further capacitors that contribute both to the overall energy of the qubit and to the coupling with the resonator. The charge coupling to the resonator is typically weak for flux qubits and will be neglected in the following. We treat any remaining capacitors by assuming an effective shunt capacitance for the normal modes of Eq. \eqref{eq:fluxqblagrange}, which we will incorporate at a later point for the precise calibration of the energy. The inner circuit may also have additional inductive components, but the energy associated with large junctions and pure geometric inductances is typically small compared to $E_J$ for small flux qubits. The inductances may, however, play a role in the structure of the mode function $u_m$ and therefore also in the coupling between the resonator and the qubit.

Due to the loop geometry, the flux node variables obey,
\begin{align}
\phi_1 + \phi_2 - \phi_3 = \Phi = \Phi_{ex} + \Delta \psi \label{eq:flux}
\end{align}
where $\Phi_{ex}$ is the externally applied flux and $\Delta \psi$ is the flux from the resonator mode fluxes across the qubit junctions. Now using Eq. \eqref{eq:flux} we can eliminate the variable $\phi_2$ and define the variables $\phi_\pm = (\phi_3 \pm \phi_1)/2$ to obtain the standard flux qubit Hamiltonian \cite{PhysRevA.80.032109, orlando1999superconducting},
\begin{align}
H_{fq} &= \frac{q_-^2}{2[(4+4\alpha)C_J + C_{S_-}]} + \frac{q_+^2}{2(4C_J + C_{S_+})}  \nonumber \\
& \quad- E_J \bigg[ 2\cos(\varphi_+) \cos(\varphi_-) + \alpha \cos\bigg(\frac{2\pi \Phi }{ \Phi_0} + 2\varphi_-\bigg) \bigg] \label{eq:Hfq}
\end{align}
with $q_i$ being the conjugate variable to $\phi_i$ and $\varphi_i = 2\pi\phi_i/\Phi_0$. The capacitances $C_{S_i}$ are the effective shunt capacitances for each mode and depend on the specific geometry of the loop. The Hamiltonian \eqref{eq:Hfq} can be diagonalized numerically and at the flux sweet-spot $\Phi_{ex} = \Phi_0/2$ the energy-splitting between the two lowest eigenstates is much smaller than $E_J$, while the third state is far away in the energy spectrum. I.e., the system is more appropriately described as a qubit than as a Kerr-nonlinear oscillator in the $\varphi_-$-mode.

Now, we exploit the fact that we can write
\begin{align}
\delta \psi = \sum_m \hat{\varphi}_m \Delta u_m
\end{align}
to write the coupling Hamiltonian (truncated to the qubit states of the circuit)
\begin{align}
H_{\phi} = \sum_m \hbar g_m \ket{0}\bra{1} (a\dag + a) + \text{h.c.}
\end{align}
with
\begin{align}
\hbar g_m = \alpha E_J \sqrt{\frac{\hbar}{2C_{\Sigma} \omega_m}} \frac{2\pi \delta u_m}{\Phi_0} \bra{0} \sin(2\varphi_- + \frac{2\pi \Phi_{ex}}{\Phi_0}) \ket{1}. \label{eq:gm}
\end{align}
Depending on the size of $\delta u_m$ this coupling may be up to $\sim 0.1 \hbar \omega_{m}$ for typical parameters \cite{PhysRevA.80.032109}.

In \mbox{M. Rehák \emph{et.al.}} \cite{rehak2014parametric} (see also Ref.~\cite{PhysRevB.91.104516}), a two qubit circuit was embedded in a transmission wave-guide resonator with the purpose to enhance the non-linear nature of the system and thus provide a very large non-linearity. While two flux qubits are naturally connected via the  
mutual inductance $M$ between the loops,  this will typically yield only a weak coupling, $H_c \propto (\phi_-^{(1)}\phi_-^{(2)})/M$, compared to the qubit energy splitting $\hbar \omega_{10} = E_{\ket{1}} - E_{\ket{0}}$. We can, however, obtain a stronger coupling by allowing the two loops to share an inductor with inductance $L_C$, which we assume much smaller than the inductance of the three flux qubit Josephson junctions so that the Hamiltonian Eq. \eqref{eq:Hfq} of the indivdual qubits is not changed. The requirement that the inductance energy of the qubits with the four junctions is the same as for with original three junctions leads to  the  condition
\begin{align}
\frac{\Phi_\Sigma^2}{2(2L_J + L_\alpha)} = \frac{(\Phi_\Sigma+\phi_C)^2}{2(2L_J + L_\alpha + L_C)}, \label{eq:Lcontraint}
\end{align}
with $L_J$ and $L_\alpha$ being the Josephson inductance corresponding to $E_J$ and $\alpha E_J$ while $\phi_C$ is the flux difference across the coupling inductor. Since the coupling inductor is shared by the two loops, the combined flux difference is given as $\phi_C = \phi_c^{(1)} \pm \phi_c^{(2)}$ where $\phi_c^{(i)}$ is the associated flux difference for the $i$th qubit loop according to Eq. \eqref{eq:Lcontraint} with the $\pm$ sign defined by the loop geometry. Finally, we obtain the coupling Hamiltonian
\begin{align}
H_c = \pm G_{12} (\ket{0}\bra{1} + \ket{1}\bra{0})\otimes(\ket{0}\bra{1} + \ket{1}\bra{0})
\end{align}
with
\begin{align}
G_{12} = \frac{8(\Phi_0/2\pi)^2}{L_C} \frac{L_C \bra{0} \phi_-^{(1)} \ket{1}}{2L_J^{(1)} + L_\alpha^{(1)}}  \frac{L_C \bra{1} \phi_-^{(2)} \ket{0}}{2L_J^{(2)} + L_\alpha^{(2)}}.
\end{align}
This result is equivalent to the expression for $G_{12} \propto I_p^{(1)} I_p^{(1)} / I_{0C}$ quoted in \cite{rehak2014parametric, PhysRevB.72.020503, PhysRevLett.96.047006} where $I_p^{(i)}$ is the persistent current in the $i$th qubit and $I_{0C}$ is the critical current of the coupling inductor, but it is modified by the magnitude of the qubit transition matrix elements. Finally, let us note that the constraint Eq. \eqref{eq:Lcontraint} is only approximate and we should have included the coupling inductor in both loops when properly deriving the flux qubit Hamiltonian. Such a full derivation gives rise to a larger flux difference across the coupling inductor and to a slightly higher coupling strength.

\begin{figure}
\includegraphics[width=0.99\linewidth]{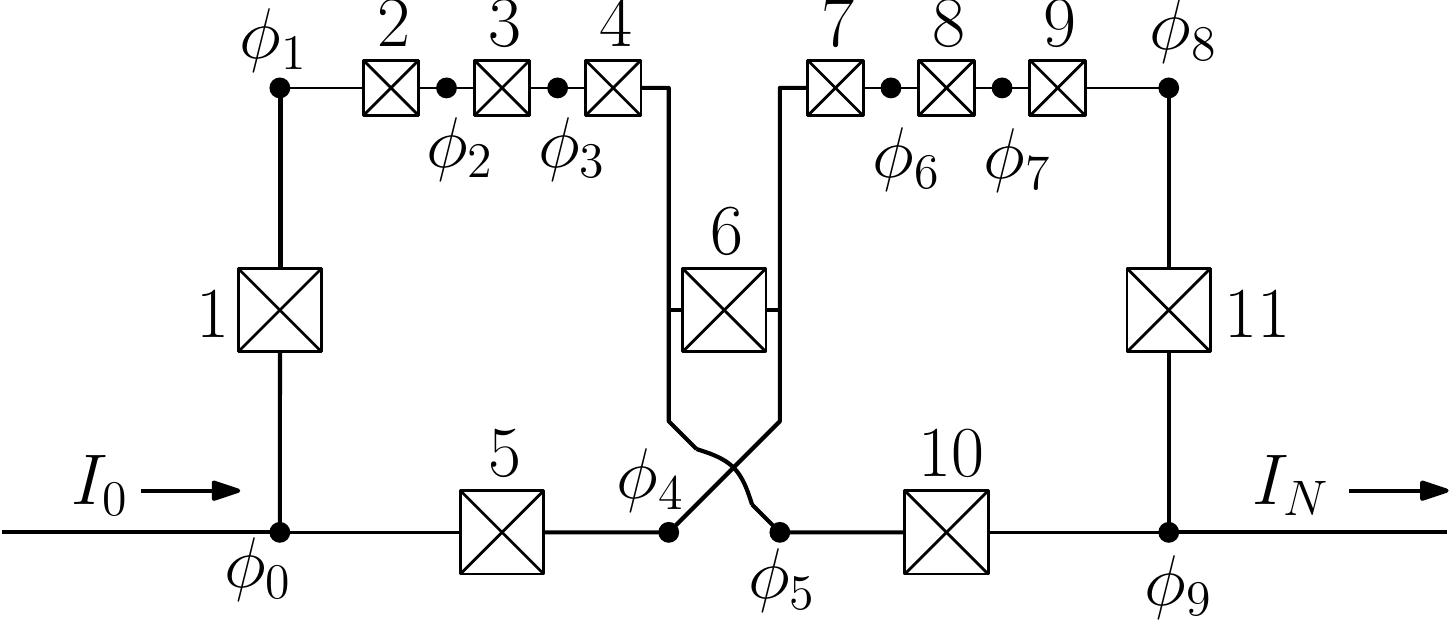}
\caption{Circuit diagram for the system with two inductively coupled flux qubits that we consider in Sec. \ref{sec:experiment}. We have 11 numbered Josephson junctions and each node flux is denoted $\phi_m$. From the left we have the incoming current $I_0$ and on the right the outgoing $I_N$ governed by the transmission-line mode.} \label{fig:circuit}
\end{figure}

\section{Analysis of a recent experiment}
\label{sec:experiment}
To demonstrate the power of our derived method we consider a recent experiment by \mbox{M. Rehák \emph{et.al.}} \cite{rehak2014parametric}  implementing the architecture shown in Fig.~\ref{fig:circuit} (same setup used in Ref.~\cite{PhysRevB.91.104516}). This experiment finds a Kerr non-linearity with $K_{33}/\omega_3 = 3\times10^{-3}$, contradicting expectations based on Eq.~\eqref{eq:Kmm} which, for typical parameters, yields a negative Kerr-nonlinearity with  $|K_{mm}/ \omega_m| \lesssim 10^{-4}$.

\begin{figure}[b]
\includegraphics[width=\linewidth]{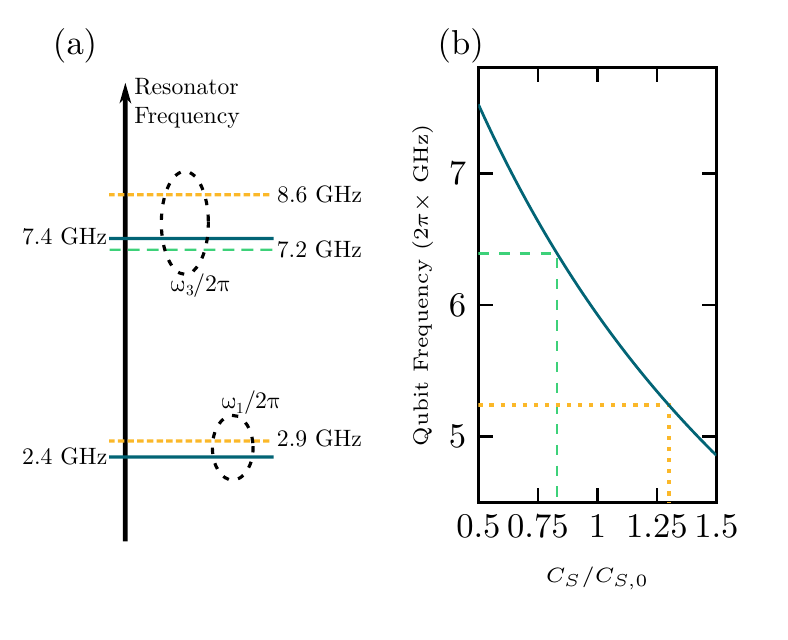}
\caption{(a) The resonator frequency. The short-dashed (orange) lines indicate the eigenfrequencies ($\pi v n/l$) of the 1st and 3rd mode (marked by the circles) for an uncoupled transmission-line, while the values shown by solid blue lines are calculated using Eq. \eqref{eq. trancendental equation for omega}. The long-dashed green line is $3\omega_1$, with $\omega_1$ being the experimentally measured frequency of the 1st mode. (b) The qubit frequency calculated by diagonalizing Eq. \eqref{eq:Hfq} is shown as function of the shunt capacitance $C_S$ in units of the value $C_{S,0}$ estimated from the experimental parameters. The dashed green and orange lines show the measured qubit frequencies for the first and second qubits and the corresponding effective shunt capacitances.} \label{fig:freq}
\end{figure}

The system studied in \cite{rehak2014parametric} is illustrated in the circuit diagram in Fig. \ref{fig:circuit} with 11 Josephson junctions placed in two loops. In each loop the three small junctions constitute a flux qubit with $C_J = 3$~fF and $E_J = 2\pi\times 60$ GHz and $\alpha = 0.5$. The two lower resonator coupling junctions, 5 and 10, have areas 10 times larger than the $\alpha$ junction and the side junction capacitances, 1, 6 and 11, are yet again 10 times larger than the ones of the resonator coupling junctions. The qubits are cross coupled by the junction, 6, using a twist technique introduced in \cite{PhysRevB.72.020503} which ensure that $H_c \sim -G_{12}$.

We can calculate the resonator frequency using Eq.~\eqref{eq. trancendental equation for omega}, presented in Fig. \ref{fig:freq}~(a), and the two qubit frequencies, shown in Fig. \ref{fig:freq}~(b). Treating  the resonator length, $l$, as an adjustable parameter we match the fundamental mode to the experimental value of $\omega_1 = 2\pi \times 2.4$ GHz, and our theory then yields the eigenfrequency of the third mode, $\omega_3 = 2\pi \times 7.4 \text{ GHz} \neq 3\omega_1$ which exactly match the experimentally measured value and is a first indication that our theory describes the experiment well.

The qubit frequency can be calculated by diagonalization of the Hamiltonian Eq.~\eqref{eq:Hfq}. If we calculate the shunt capacitance, $C_{S,0}$, from the serial line of capacitances of the three large junctions 1,5 and 6 (equivalent to 6, 10 and 11) we obtain a qubit frequency of $\omega_{10} = 2\pi\times 5.9$ GHz, different from the measured values $\omega_{10}^{(1)} = 2\pi \times 6.39$ GHz and $\omega_{10}^{(2)} = 2\pi \times 5.28$ \cite{PhysRevB.91.104516}. By allowing adjustment of the shunt capacitances, $C_S$, for both qubits we can, however, reproduce the correct qubit frequencies, see Fig. \ref{fig:freq}~(b).

 After applying the parameter adjustments to match the frequency of each component, we can proceed and determine the non-linear properties of the system. From Eq. \eqref{eq:Kmm} we can thus directly calculate the Kerr constant for the third mode, which was probed in the experiment. We find a value of $K_{33} / \omega_3 = -3 \times 10^{-5}$, which is much smaller than the measured result and of the opposite sign. We know, however, that this expression does not properly take into account the Rabi splitting of the eigenmodes by the coupling to the qubit degrees of freedom. To evaluate the effects of the non-linear terms we must diagonalize the Hamiltonian,
\begin{align} \label{eq:Hfin}
H_3 &= \hbar \omega_3 a\dag a + \frac{\hbar K_{33}}{2} a\dag a a\dag a
+ \frac{\hbar \omega_{10}^{(1)}}{2} \sigma_z^{(1)} + \frac{\hbar \omega_{10}^{(2)}}{2} \sigma_z^{(2)} \nonumber \\
&\quad + \hbar g_3^{(1)} \sigma_x^{(1)} (a\dag + a)
 + \hbar g_3^{(2)} \sigma_x^{(2)} (a\dag + a) \nonumber \\
&\quad -G_{12}\, \sigma_x^{(1)}\sigma_x^{(2)},
\end{align}
where $\sigma_x^{(i)} = (\ket{0^{(i)}}\bra{1^{(i)}} + \ket{1^{(i)}}\bra{0^{(i)}})$, and only then extract the effective Kerr constant, $\tilde{K}_{33}$. 

We calculate the coupling strength $g_3^{(i)}$ using Eq. \eqref{eq:gm} where the flux differences $\delta u_3^{(1)}$ and $\delta u_3^{(2)}$ come from $\phi_5 - \phi_1$ and $\phi_4 - \phi_8$ of Fig. \ref{fig:circuit}, respectively. The qubit-qubit coupling, $G_{12}$, is derived from the experimental design parameters and is negative due to the twisted coupling between $\phi_5$ and $\phi_4$. With these values and a numerical diagonalization of Eq. \eqref{eq:Hfin}, we obtain $\tilde{K}_{33} / \omega_3 = 2.1 \times 10^{-3}$, which is, indeed, close to the experimentally obtained value. In comparison, for a single flux qubit (assuming $g_3^{(2)} = G_{12} = 0$) we find $\tilde{K}_{33}^{(1)} / \omega_3 = 3.28 \times 10^{-4}$.

\begin{figure}[t]
\includegraphics[width=8cm]{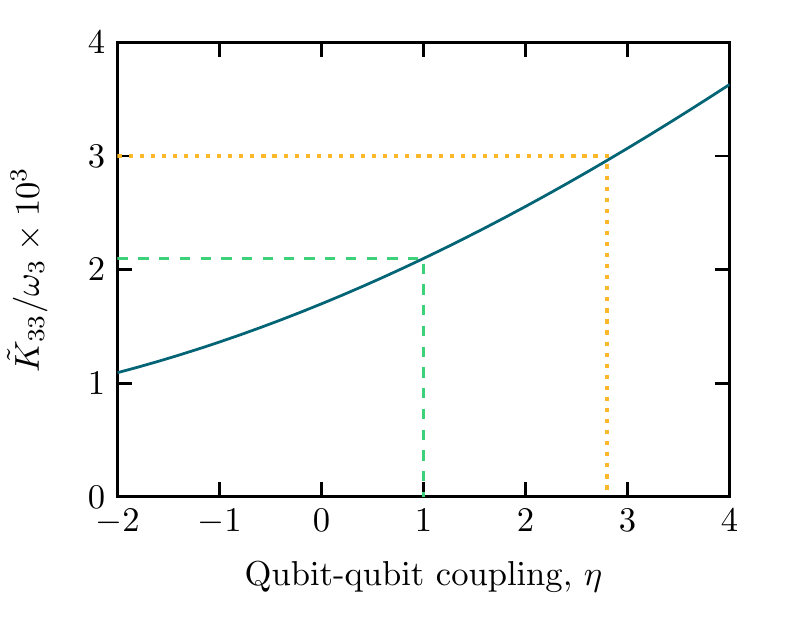}
\caption{The effective Kerr constant $\tilde{K}_{33}^{(1)} / \omega_3$ as a function of the relative qubit-qubit coupling is shown as the solid blue line. The dashed green line marks the value obtained by directly using the experimental design parameters, while the orange dotted line marks the experimentally measure Kerr constant.} \label{fig:K}
\end{figure}

As we are expecting to underestimate the value of the qubit-qubit coupling, $G_{12}$, we have investigated the role of its value in the determination of $\tilde{K}_{33}^{(1)} / \omega_3$, by scaling it with a constant factor $G_{12} \rightarrow \eta\, G_{12}$  without changing the other parameters of the problem. Fig. \ref{fig:K} shows how the resulting effective Kerr coefficient depends on $\eta$. We observe that an increase in the qubit-qubit coupling leads to an increased Kerr constant but also that, even if the qubits do not interact directly, we still obtain a sizeable Kerr constant.

As also demonstrated in \cite{rehak2014parametric}, the external flux can be tuned in-situ and, thus, yielding an effective tunable Kerr constant of the combined system. Most significantly, the external flux can be put at $\Phi_x = 0$, where the qubit mode is negligible and, thus, the non-linearity is accurately predicted by Eq. \eqref{eq:Kmm}. Thus we can both tune the sign and magnitude of the non-linearity.

\section{Conclusion and outlook}
\label{sec:conc}

In this work we have presented a general framework to obtain an effective quantum description of a discrete circuit embedded in a transmission-line. Our approach treats the coupled systems by separately driving the circuit input and output nodes with current variables, which in turn, are forced to match the wave equation and boundary conditions for the transmission-line. This approach extends the successful Lagrangian formalism for quantization of circuit systems, which has been vastly successful in the field of circuit QED. 

We illustrated the use of our method by an application to flux qubits, which allowed us to establish a full quantum model for a recent experiment performed by \mbox{M. Rahák \emph{et.al.} \cite{rehak2014parametric}}. By numerical diagonalization of the few mode quantum Hamiltonian, we were able to calculate the effective Kerr coefficient of the transmission-line in good agreement with the experimentally obtained value.

Our method may be directly applicable to any design using inductors, capacitors and Josephson junction embedded in a transmission-line resonator, and we expect that generalization is also possible to travelling wave systems, where it may be used to to describe the recently developed Josephson travelling wave parametric amplifiers \cite{macklin2015near, white2015traveling, grimsmo2016engineering}.

\section*{Acknowledgements}
 The authors are grateful to J. Bourrassa for discussions and valuable feedback and to M.~Rehák, P.~Neilinger and M.~Grajcar for discussions and for providing the parameters used in Sec. \ref{sec:experiment}. This works was supported by the Villum Foundation, and CKA acknowledges support from the Danish Ministry of Higher Education and Science.

\bibliography{reference}
\end{document}